\pdfoutput=1
\documentclass[%
  aip,
  amsmath,amssymb,
  reprint,
  groupedaddress
]{revtex4-1}

\usepackage{graphicx}
\DeclareGraphicsExtensions{.pdf,.png,.jpg,.jpeg}
\usepackage{dcolumn}
\usepackage{bm}

\usepackage[utf8]{inputenc}
\usepackage[T1]{fontenc}
\usepackage{mathptmx}

\usepackage[english]{babel}
\usepackage{microtype}

\usepackage{physics}
\usepackage{mathtools}
\usepackage{siunitx}
\usepackage{hyperref}
\usepackage{cleveref}
\usepackage{enumitem}
\usepackage{booktabs}
\usepackage{caption}
\usepackage{subcaption}
\usepackage{float}
\hypersetup{
  colorlinks=true,
  linkcolor=black,
  citecolor=black,
  urlcolor=black
}

\makeatletter
\def\frontmatter@title@below{\addvspace{0.75\baselineskip}}
\makeatother

\newcommand{\kvec}{\bm{k}}
\newcommand{\ivec}{\bm{i}}
\newcommand{\jvec}{\bm{j}}
\newcommand{\eR}{\bm{e}_\rho}
\newcommand{\eP}{\bm{e}_\varphi}
\newcommand{\rr}{\bm{r}}
\newcommand{\vv}{\bm{v}}
\newcommand{\aaa}{\bm{a}}
\newcommand{\FF}{\bm{F}}
\newcommand{\MM}{\bm{M}}
\newcommand{\NN}{\bm{N}}
\newcommand{\Tf}{\bm{T}}
\newcommand{\what}[1]{\widehat{#1}}

\begin{document}

\title{A tabletop demonstration of distributed friction: the spinning wine glass}

\author{Rubén Canora}
\email{ruben.canora@gmail.com}
\date{\today}

\begin{abstract}
When a wine glass is dragged on a table along a circular path, a spontaneous rotation about its vertical axis can develop even if the applied hand force does not directly introduce a yaw torque. This document provides a structured formal derivation of the governing equations that are responsible for this behavior.

The analysis shows that the mechanism responsible for this effect is the redistribution of pressure onto the table when applying the force with your hand. This causes an uneven frictional force distribution which exerts a torque on the glass, causing it to spin.
\end{abstract}
\maketitle

% -----------------------------------------------------------
\section{\label{sec:introduction}Introduction}

Drag a wine glass on a tabletop so that its center follows a circle. If the force is applied at the stem, the glass begins to yaw in the opposite direction of the orbital motion, even though the hand appears to impose only translation. What produces this spontaneous spin?

The appeal of this phenomenon is that it is easy to reproduce at home. Take a glass of wine and repeat the following experiment twice, first exerting the force close to the center of mass (CM) and second at a distance from the CM closer to the base. Make an orbital motion with the glass, same speed, same radius (without impeding yaw rotation with your hands). The first typically shows little or no yaw, while the second produces a clear self-rotation in the opposite direction of the movement. 

This experiment already points to the essential physics: the table does not identically react to each point in the base, but through a distributed contact over the annular rim of the base. The normal load is therefore not uniform around the ring, and whenever your applied force is vertically offset from the CM it produces a rotational moment that shifts where the glass ``presses'' on the table. This pressure redistribution makes frictional forces stronger on one side than the other, breaking symmetry and generating a net torque about the vertical axis. Related issues arise broadly in contact mechanics and tribology, where the real area of contact and the distribution of normal stress are central to the resulting frictional force \cite{bowden1950friction,greenwood1966contact,johnson1985contact,persson2000sliding}.

% In AIP reprint (two-column), a near-full-width figure is best as figure*.
\begin{figure*}[t]
    \centering
    \includegraphics[width=0.85\textwidth]{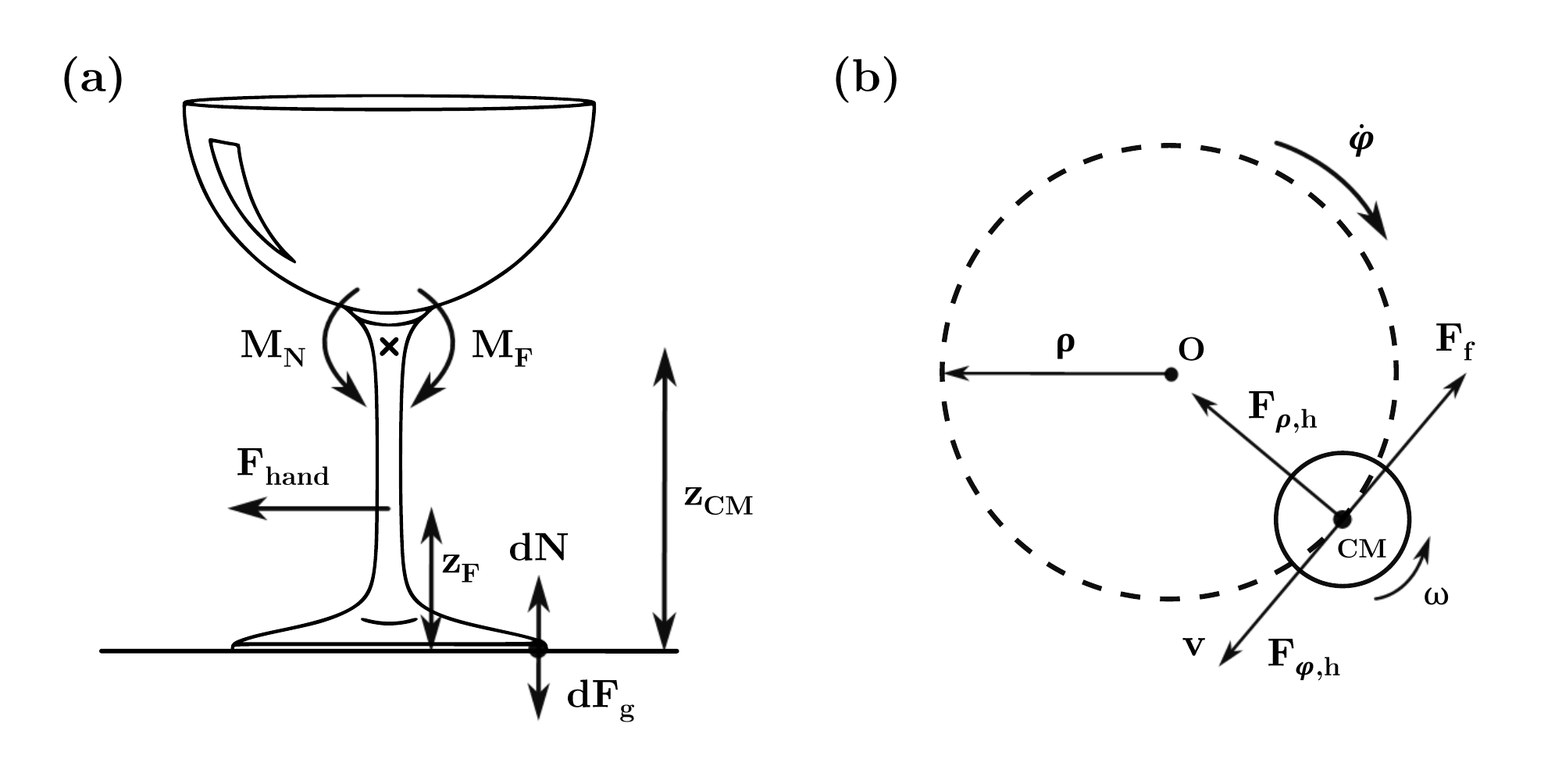}
    \caption{\textbf{Mechanical Configuration of the System.} (a) Side View: The hand force $\mathbf{F}_{\text{hand}}$ is applied at height $z_F$, offset from the Center of Mass (CM) at height $z_{CM}$. This vertical misalignment generates an overturning moment about the CM which is balanced by the moment about the CM produced by the redistributed normal reaction, $M_N$. (b) Top-Down View: The glass follows a circular path of radius $\rho$ around center $O$. The applied force has radial ($F_{\rho,h}$) and tangential ($F_{\varphi,h}$) components to maintain the orbital trajectory against friction $\mathbf{F}_f$.}
    \label{fig:problem_diagram}
\end{figure*}

The problem statement is best understood through the geometry illustrated in Figure \ref{fig:problem_diagram}. The glass is constrained to move with tangential velocity $\mathbf{v}$ along a circle of radius $\rho$. To maintain this trajectory, the force of the external hand must provide the necessary centripetal acceleration ($F_{\rho,h}$) to overcome the tangential friction ($F_{\varphi,h}$). In what follows, we show how the same vertical offset that produces the overturning moment also sets the leading yaw torque through the redistributed contact forces.

\subsection{\label{subsec:educational_value}Educational value and intended audience}

The purpose of this article is not only to explain a curious tabletop effect, but also to provide a simple teaching example for undergraduate mechanics. The demonstration is suitable for students who have already encountered Newton's laws, torque, rigid-body rotation and Coulomb friction, but who may still be accustomed to treating friction as a single resultant force acting at a point.

The spinning-glass experiment is useful pedagogically because it separates several ideas that are often conflated in introductory treatments. First, it shows that a horizontal force can produce an overturning moment without directly producing a yaw torque. Second, it illustrates that the normal reaction of a table is not, in general, a single force applied at a fixed point, but the resultant of a distributed contact pressure. Third, it shows that frictional torques may arise from the spatial distribution of friction forces, even when the applied hand force appears to impose only translation.

The main learning objectives are therefore:
\begin{enumerate}[label=(\roman*)]
    \item to distinguish the net friction force from the distributed friction field;
    \item to identify how an overturning moment redistributes the normal load over an extended contact;
    \item to derive a simple first-harmonic model for the pressure distribution around an annular base;
    \item to obtain a leading-order expression for the yaw torque generated by asymmetric friction;
    \item to compare an analytically tractable approximation with a numerical finite-slip model.
\end{enumerate}

For these reasons, the example can be used either as a lecture demonstration, a short computational project, or a laboratory activity based on video tracking. It also provides a useful bridge between the idealized point-contact friction model commonly introduced in elementary mechanics and the more realistic distributed-contact viewpoint used in tribology and contact mechanics \cite{bowden1950friction,johnson1985contact,persson2000sliding}.

% ===========================================================
\subsection{\label{sec:demonstration}Tabletop demonstration and video material}

The effect can be observed in a simple tabletop demonstration. In Supplementary Video 1, the glass is first moved along a counterclockwise circular path, while its yaw rotation develops in the opposite, clockwise direction. The motion is then reversed: when the imposed circular path is clockwise, the glass yaws counterclockwise. This reversal shows that the self-rotation is not caused by an accidental twist of the hand, but by the frictional torque generated by the redistributed normal load at the base.

\begin{center}
\href{https://youtube.com/shorts/EG6ZMKajgB4?si=jG7rwtDxJuLVwSPu}{Supplementary Video 1: reversal of yaw direction in the spinning-glass demonstration}.
\end{center}

This qualitative observation is consistent with the model developed below: changing the sign of the orbital angular velocity reverses the direction of the friction-induced yaw torque. Moreover, one can easily test that the further the application of the force from the center of mass, the bigger the resulting yaw angular velocity.

% ===========================================================
\section{\label{sec:kinematics}Kinematics and Reference Frames}

Let $(\ivec,\jvec,\kvec)$ be a fixed inertial basis where the tabletop is the plane $z=0$ and $\kvec$ is the vertical unit vector. For a CM position $\rr(t)$ on the table, we use polar unit vectors $\eR, \eP$. Imposing a constant radius circular path ($\rho(t) \equiv \rho$) and a constant orbital angular speed ($\dot{\varphi}(t)\equiv \Omega$), the kinematics simplify to:
\begin{align}
\rr &= \rho\,\eR, \\
\vv &= \rho\Omega\,\eP, \\
\aaa &= -\rho\Omega^2\,\eR.
\end{align}
Thus, the acceleration is purely centripetal. In particular, since $\dot{\varphi}$ is constant we have $\ddot{\varphi}=0$, so there is no tangential acceleration component.

On the other hand, the glass rotates about the vertical axis through the CM with angular velocity $\bm{\omega}(t) = \omega(t)\,\kvec$ and angular acceleration $\dot{\bm{\omega}}=\dot{\omega}\,\kvec$. We denote the moment of inertia about this axis as $I_z$.

% ===========================================================
\section{\label{sec:forces}Forces, moments, and equations of motion}

\subsection{\label{subsec:external_forces}External forces}
The forces acting on the glass are:
\begin{enumerate}[label=(\roman*)]
\item Weight: $-mg\,\kvec$.
\item Normal reaction from the table, distributed along the contact ring.
\item Friction force from the table, distributed along the contact ring.
\item The hand-applied force $\FF$ (assumed horizontal, i.e.\ in the plane $xy$).
\end{enumerate}

\subsection{\label{subsec:overturning_moment}Overturning moment due to the hand force}
Let the hand force be applied at height $z_F$ above the table, while the CM is at height $z_{CM}$. Define the vertical offset of the force application point relative to the CM:
\begin{equation}
\Delta z \equiv z_F - z_{CM}.
\end{equation}

Thus, the position vector from CM to the force application point is purely vertical:
\begin{equation}
\bm{r}_{F/CM} = \Delta z\,\kvec.
\end{equation}

The moment (torque) of a force about the CM is defined by
\begin{equation}
\MM = \bm{r}\times \FF,
\end{equation}
where $\bm{r}$ is the vector from CM to the point of application. Therefore the moment produced by the hand force about the CM is:
\begin{equation}
\boxed{
\MM_h = \bm{r}_{F/CM}\times \FF = \Delta z\,\kvec \times \FF.
}
\end{equation}

Since $\FF$ is horizontal and $\kvec$ is vertical, $\kvec\times\FF$ lies in the horizontal plane. Therefore $\MM_h$ is a horizontal moment: it tends to tilt the glass, not yaw it directly. In particular,
\begin{equation}
\MM_h\cdot \kvec = 0.
\end{equation}

\subsection{\label{subsec:newton_euler}Newton's and Euler's equations of motion}
The translational equation of motion of the CM (in the horizontal plane) is
\begin{equation}
\boxed{
m\aaa = \FF + \Tf,
}
\end{equation}
where $\Tf$ is the total friction force from the table (which results from the distributed friction along the ring). \\

For rotation about the vertical axis through the CM, Euler's equation reduces to
\begin{equation}
\boxed{
I_z \dot{\omega} = \tau_z,
}
\end{equation}
where $\tau_z$ is the net torque about the vertical axis produced by all contact forces. 

% ===========================================================
\section{\label{sec:contact}Contact modelling: annular ring, normal load distribution and tilt equilibrium}

\subsection{\label{subsec:ring_geometry}Geometry of the contact ring}
The ring contact is modeled as a circumference of radius $R$ in the horizontal plane, centered at the vertical projection of the CM (assuming no/small tilts).

A point on the ring is parameterized by an angular coordinate $\theta$:
\begin{equation}
\boxed{
\bm{r}_c(\theta) = R\cos\theta\,\ivec + R\sin\theta\,\jvec.
}
\end{equation}

\subsection{\label{subsec:normal_distribution}Normal force distribution}
Let $\lambda(\theta)$ be the normal load per unit angle (N/rad) along the ring. Then the normal force element is
\begin{equation}
\boxed{
d\NN(\theta) = \lambda(\theta)\,\kvec\, d\theta.
}
\end{equation}

The total normal force is obtained by integration:
\begin{equation}
\boxed{
N = \int_0^{2\pi}\lambda(\theta)\,d\theta.
}
\end{equation}
In the absence of vertical acceleration,
\begin{equation}
N \approx mg.
\end{equation}

\subsection{\label{subsec:normal_moment}Moment of the distributed normal force about the CM}
Each normal force element produces a moment about the CM:
\begin{equation}
d\MM_N(\theta) = \bm{r}_c(\theta)\times d\NN(\theta).
\end{equation}
Substituting $\bm{r}_c=(x,y,0)$ and $d\NN=(0,0,\lambda d\theta)$:
\begin{align}
d\MM_N &= (x\ivec+y\jvec)\times(\lambda\kvec\,d\theta) \\
&= \lambda d\theta \Big(x(\ivec\times \kvec) + y(\jvec\times\kvec)\Big).
\end{align}
Using $x(\theta)=R\cos\theta$ and $y(\theta)=R\sin\theta$, as well as $\ivec\times\kvec=-\jvec$ and $\jvec\times\kvec=\ivec$:
\begin{equation}
\boxed{
d\MM_N(\theta) = \big(R\sin\theta\,\lambda(\theta)\,\ivec - R\cos\theta\,\lambda(\theta)\,\jvec\big)\,d\theta.
}
\end{equation}
Therefore the horizontal components of the normal-induced moment are:
\begin{equation}
\boxed{
\begin{aligned}
M_{N,x} &= \int_0^{2\pi} R\sin\theta\,\lambda(\theta)\,d\theta,\\
M_{N,y} &= -\int_0^{2\pi} R\cos\theta\,\lambda(\theta)\,d\theta.
\end{aligned}
}
\end{equation}

\subsection{\label{subsec:tilt_equilibrium}Quasi-static tilt equilibrium}
We assume the glass does not appreciably accelerate in roll/pitch (its tilt adjusts rapidly so that overturning moments balance). Under this condition,
\begin{equation}
\sum \MM_{\text{horizontal}} \approx \bm{0}.
\end{equation}
The only significant horizontal moments in the minimal model are:
\begin{itemize}
\item $\MM_h$ from the hand force.
\item $\MM_N$ from the shifted distribution of the normal load.
\end{itemize}
Thus the equilibrium reads:
\begin{equation}
\boxed{
\MM_h + \MM_N = \bm{0}
\quad\Longrightarrow\quad
\MM_N = -\MM_h.
}
\end{equation}

% ===========================================================
\section{\label{sec:pressure_distribution}Effect of the force on pressure distribution}

\subsection{\label{subsec:first_harmonic}First-harmonic ansatz}
A small tilt produces a smooth increase of contact pressure on one side of the ring and a smooth decrease on the opposite side. In order to model this behavior, the choice of a first-harmonic form can be justified from equilibrium constraints and minimality.

For a rigid contact modeled as a ring, the normal traction $\lambda(\theta)$ must satisfy two requirements: (i) the total normal force is fixed,
\[
\int_0^{2\pi}\lambda(\theta)\,d\theta = N,
\]
and (ii) the redistributed normal load must produce the required in-plane moment about the center of mass, and its components have to be determined by the first angular moments
\[
\int_0^{2\pi}\lambda(\theta)\cos\theta\,d\theta,\qquad
\int_0^{2\pi}\lambda(\theta)\sin\theta\,d\theta.
\]
These are precisely the Fourier coefficients of the first angular harmonic. Consequently, the decomposition
\[
\lambda(\theta)=\frac{N}{2\pi}+\sum_{n\ge1}\big(a_n\cos n\theta+b_n\sin n\theta\big)
\]
shows that the constant term alone enforces the correct total normal force, while the pair $(a_1,b_1)$ is the lowest-order contribution that can represent an in-plane shifting of the resultant normal load. All higher harmonics ($n\ge2$) do not interfere in these global balances and therefore only refine the local shape of $\lambda(\theta)$ without improving the closure for the net force and moment at this order.

For this reason, we use the Fourier series at first order to define the minimal ansatz
\begin{equation}
\boxed{
\lambda(\theta)=\frac{N}{2\pi}+A\cos\theta+B\sin\theta,
}
\end{equation}
that will optimally define the change in pressure distribution.

\subsection{\label{subsec:normal_moment_AB}Computing $M_{N,x}$ and $M_{N,y}$ in terms of $A,B$}
For the x-direction:
\begin{align}
M_{N,x} &= \int_0^{2\pi} \big(R\sin\theta\big)\left(\frac{N}{2\pi}+A\cos\theta+B\sin\theta\right)\,d\theta.
\end{align}
The integrals
\[
\int_0^{2\pi}\sin\theta\,d\theta=0,\quad
\int_0^{2\pi}\sin\theta\cos\theta\,d\theta=0,\quad
\int_0^{2\pi}\sin^2\theta\,d\theta=\pi
\]
imply
\begin{equation}
\boxed{
M_{N,x}=\pi R B.
}
\end{equation}

For the y-direction:
\begin{align}
M_{N,y} &= -\int_0^{2\pi} \big(R\cos\theta\big)\left(\frac{N}{2\pi}+A\cos\theta+B\sin\theta\right)\,d\theta.
\end{align}
Using
\[
\int_0^{2\pi}\cos\theta\,d\theta=0,\quad
\int_0^{2\pi}\cos\theta\sin\theta\,d\theta=0,\quad
\int_0^{2\pi}\cos^2\theta\,d\theta=\pi,
\]
we obtain
\begin{equation}
\boxed{
M_{N,y}=-\pi R A.
}
\end{equation}\\

The vector condition $\MM_N=-\MM_h$ implies:
\begin{equation}
M_{N,x}=-M_{h,x},\qquad M_{N,y}=-M_{h,y}.
\end{equation}
Substituting $M_{N,x}=\pi RB$ and $M_{N,y}=-\pi RA$ gives
\begin{equation}
\pi RB = -M_{h,x},
\qquad
-\pi RA = -M_{h,y}.
\end{equation}
Therefore
\begin{equation}
\boxed{
A=\frac{M_{h,y}}{\pi R},
\qquad
B=-\frac{M_{h,x}}{\pi R}.
}
\end{equation}

\subsection{\label{subsec:final_lambda}Final expression for the pressure distribution}
We can now write $\lambda(\theta)$ in terms of the hand-induced overturning moment:
\begin{equation}
\boxed{
\lambda(\theta)=\frac{N}{2\pi}
+\frac{M_{h,y}}{\pi R}\cos\theta
-\frac{M_{h,x}}{\pi R}\sin\theta.
}
\end{equation}

Physically, the normal load cannot be negative at any point, as all points in the ring remain attached to the surface:
\begin{equation}
\lambda(\theta)\ge 0 \quad \text{for all } \theta.
\end{equation}

% ===========================================================
\section{\label{sec:friction}Friction modelling and the origin of the yaw torque}

\subsection{\label{subsec:coulomb}Local Coulomb friction law on the ring}
At each ring element, the friction force magnitude is proportional to the local normal load:
\[
\abs{d\bm{F}_f} = \mu\, dN = \mu\,\lambda(\theta)\,d\theta,
\]
where $\mu$ is the coefficient of kinetic friction. The friction direction is opposite the local slip velocity of that ring point relative to the table.

\subsection{\label{subsec:slip_velocity}Local slip velocity at the ring}
A point on the ring has velocity composed of:
\begin{itemize}
\item Translation of the CM: $\vv$.
\item Rotation about the CM due to yaw: $\bm{\omega}\times \bm{r}_c(\theta)$.
\end{itemize}
Hence the local velocity of the contact point relative to the table is
\begin{equation}
\boxed{
\bm{v}_{\mathrm{rel}}(\theta) = \vv + \bm{\omega}\times \bm{r}_c(\theta).
}
\end{equation}
In our yaw-only model, $\bm{\omega}=\omega\kvec$.

The local Coulomb friction element is therefore
\begin{equation}
\boxed{
d\bm{F}_f(\theta) = -\mu\,\lambda(\theta)\,\frac{\bm{v}_{\mathrm{rel}}(\theta)}{\norm{\bm{v}_{\mathrm{rel}}(\theta)}}\,d\theta.
}
\end{equation}

Being the total friction acting on the body:
\begin{equation}
\Tf = \int_0^{2\pi} d\bm{F}_f(\theta).
\end{equation}

\subsection{\label{subsec:tau_definition}Yaw torque: definition and physical meaning}
\vspace{ -0.5cm}
\begin{figure}[H]
  \centering
  \includegraphics[width=0.7\linewidth]{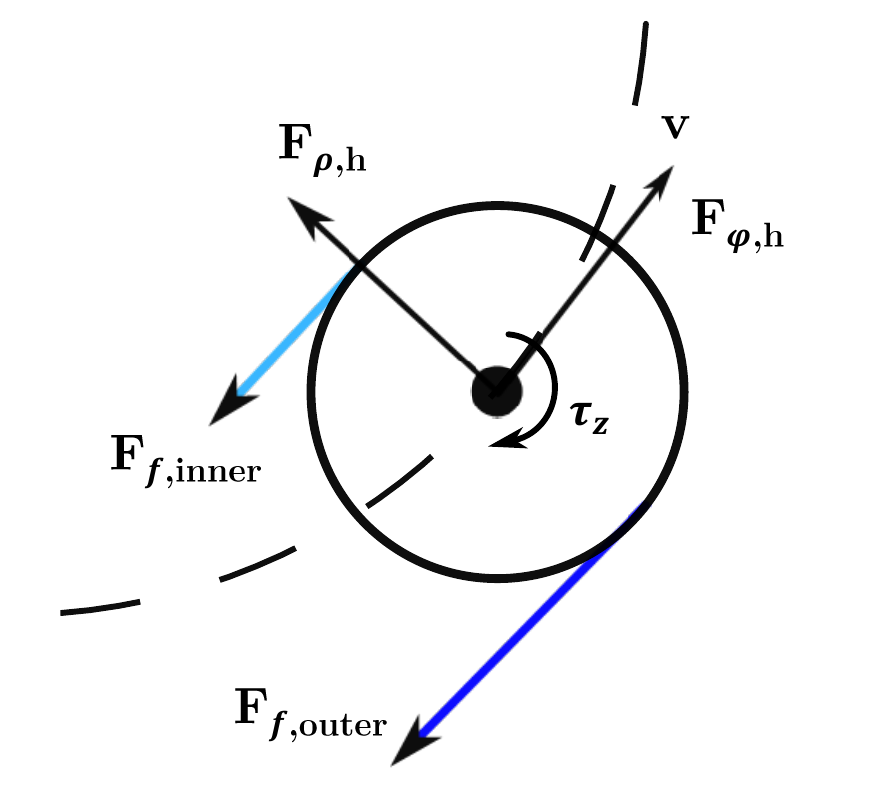}
  \caption{%
  \textbf{Yaw torque from asymmetric friction.}
  A non-uniform normal load around the contact ring produces unequal friction forces at opposite points, breaking symmetry and generating a net yaw moment $\tau_z$ about the CM.}
  \label{fig:yaw_torque_asym_friction}
\end{figure}

The moment of an elemental friction force about the CM is
\[
d\MM_f(\theta) = \bm{r}_c(\theta)\times d\bm{F}_f(\theta).
\]
The yaw torque is the vertical component of the total friction moment:
\begin{equation}
\boxed{
\tau_z = \left(\int_0^{2\pi} \bm{r}_c(\theta)\times d\bm{F}_f(\theta)\right)\cdot \kvec.
}
\end{equation}
As shown in Fig.~\ref{fig:yaw_torque_asym_friction}, for a force applied below the CM, pressure redistribution will cause a larger friction force in the outer part, causing it to rotate.

\section{\label{sec:translation_dominated}Analytically tractable regime: translation-dominated slip}

A key simplification occurs when yaw is initially small compared with translation, so that the slip direction is approximately uniform along the ring:
\begin{equation}
\bm{v}_{\mathrm{rel}}(\theta) \approx \vv
\quad\Longrightarrow\quad
\frac{\bm{v}_{\mathrm{rel}}(\theta)}{\norm{\bm{v}_{\mathrm{rel}}(\theta)}} \approx \frac{\vv}{\norm{\vv}} \equiv \what{\vv}.
\end{equation}
Then
\begin{equation}
\boxed{
d\bm{F}_f(\theta) \approx -\mu\,\lambda(\theta)\,\what{\vv}\,d\theta.
}
\end{equation}

\subsection{\label{subsec:tau_closed_form}Closed-form evaluation of $\tau_z$}
We now compute $\tau_z$ explicitly under the same approximation. For a constant-radius circular path, the CM velocity is tangent to the orbit, hence
\begin{equation}
\what{\vv}=\eP.
\end{equation}

The contact ring (relative to the CM) is parameterized as before by an angle $\theta$ measured from the fixed $\ivec$ axis:
\begin{equation}
\bm{r}_c(\theta)=R\cos\theta\,\ivec+R\sin\theta\,\jvec,
\end{equation}
and the elemental normal force is $d\bm{N}(\theta)=\lambda(\theta)\,\kvec\,d\theta$. In this regime, the slip direction is approximately set by the CM translation, so the elemental friction force is
\begin{equation}
d\bm{F}_f(\theta)\approx -\mu\,\lambda(\theta)\,\what{\vv}\,d\theta
= -\mu\,\lambda(\theta)\,\eP\,d\theta.
\end{equation}

The elemental friction moment about the CM is
\[
d\MM_f(\theta)=\bm{r}_c(\theta)\times d\bm{F}_f(\theta),
\]
and the yaw torque is the vertical component of the total friction moment:
\begin{align}
\tau_z
&=\left(\int_0^{2\pi}\bm{r}_c(\theta)\times d\bm{F}_f(\theta)\right)\cdot\kvec \nonumber\\
&\approx -\mu\int_0^{2\pi}\lambda(\theta)\,\big(\bm{r}_c(\theta)\times\eP\big)\cdot\kvec\,d\theta.
\end{align}
Using the scalar triple product,
\[
\big(\bm{r}_c\times\eP\big)\cdot\kvec
=\bm{r}_c\cdot\big(\eP\times\kvec\big),
\]
and the identity $\eP\times\kvec=\eR$, we obtain
\begin{equation}
\tau_z \approx -\mu\int_0^{2\pi}\lambda(\theta)\,\bm{r}_c(\theta)\cdot\eR\,d\theta.
\end{equation}

Let $\varphi(t)$ be the orbital angle of the CM, so that $\eR=\cos\varphi\,\ivec+\sin\varphi\,\jvec$ and $\eP=-\sin\varphi\,\ivec+\cos\varphi\,\jvec$. Then
\[
\bm{r}_c(\theta)\cdot\eR
=R\cos\theta\cos\varphi+R\sin\theta\sin\varphi
=R\cos(\theta-\varphi),
\]
and hence
\begin{equation}
\boxed{
\tau_z = -\mu R\int_0^{2\pi}\cos(\theta-\varphi)\,\lambda(\theta)\,d\theta.
}
\end{equation}

Finally, substitute $\lambda(\theta)=\frac{N}{2\pi}+A\cos\theta+B\sin\theta$ and use
\begin{equation}
\begin{aligned}
\int_0^{2\pi}\cos(\theta-\varphi)\,d\theta &= 0,\\
\int_0^{2\pi}\cos(\theta-\varphi)\cos\theta\,d\theta &= \pi\cos\varphi,\\
\int_0^{2\pi}\cos(\theta-\varphi)\sin\theta\,d\theta &= \pi\sin\varphi.
\end{aligned}
\end{equation}

This yields
\begin{equation}
\boxed{
\tau_z = -\mu\pi R\big(A\cos\varphi+B\sin\varphi\big).
}
\end{equation}
Using the earlier relations between the first-harmonic coefficients and the hand-induced overturning moment,
\[
A=\frac{M_{h,y}}{\pi R},\qquad B=-\frac{M_{h,x}}{\pi R},
\]
we substitute into the previous expression to obtain
\[
\tau_z=-\mu\big(M_{h,y}\cos\varphi - M_{h,x}\sin\varphi\big).
\]
Noting that the tangential unit vector is $\eP=-\sin\varphi\,\ivec+\cos\varphi\,\jvec$ and that
$\MM_h=M_{h,x}\,\ivec+M_{h,y}\,\jvec$, we recognize
\[
\MM_h\cdot \eP = M_{h,y}\cos\varphi - M_{h,x}\sin\varphi.
\]
Therefore, we obtain the compact invariant statement
\begin{equation}
\boxed{
\tau_z=-\mu\,(\MM_h\cdot \eP)
= -\mu\,(\MM_h\cdot \what{\vv}).
}
\end{equation}

\textbf{Interpretation.}
In the translation-dominated regime, friction is approximately opposite to the tangential direction $\eP$, so only the component of the horizontal overturning moment $\MM_h$ parallel to the motion contributes to the yaw torque. If $\MM_h=\bm{0}$ (e.g.\ $\Delta z=0$), then $\tau_z=0$.\\

\subsection{\label{subsec:polar_decomp}Force decomposition and circular-motion consequence}
By introducing the force exerted by the hand
\begin{equation}
\FF = F_\rho\,\eR + F_\varphi\,\eP.
\end{equation}
into the previously stated $\MM_h$:
\[
\MM_h = \Delta z\,\kvec\times \FF.
\]
one obtains the final form for the overturning moment in terms of the force:
\begin{equation}
\boxed{
\MM_h = \Delta z\left(F_\rho\,\eP - F_\varphi\,\eR\right).
}
\end{equation}

Moreover, for circular motion with constant radius, the velocity is
\[
\vv = \rho\dot{\varphi}\,\eP.
\]
The unit vector along the direction of motion is
\begin{equation}
\what{\vv}=\frac{\vv}{\norm{\vv}} = \frac{\rho\dot{\varphi}\,\eP}{\rho|\dot{\varphi}|}=\frac{\dot{\varphi}}{|\dot{\varphi}|}\,\eP.
\end{equation}
In the remainder of this section, we assume the circle is traversed in the direction of increasing $\varphi$, so that $\dot{\varphi}>0$. Then $\what{\vv}=\eP$. If the motion is in the opposite direction, the final torque direction reverses consistently; the vector formula $\tau_z=-\mu(\MM_h\cdot \what{\vv})$ remains valid.

Under this condition we get the following:
\begin{equation}
\begin{aligned}
\MM_h\cdot\what{\vv} = \MM_h\cdot \eP = \Delta z\left(F_\rho\,\eP\cdot\eP - F_\varphi\,\eR\cdot\eP\right)=\Delta z\,F_\rho.
\end{aligned}
\end{equation}
Therefore, the yaw torque becomes
\begin{equation}
\boxed{
\tau_z = -\mu\,\Delta z\,F_\rho
\qquad(\dot{\varphi}>0\ \text{convention}).
}
\end{equation}

\textbf{Key conclusion.}
In circular motion, the yaw torque generated by this mechanism is controlled by the radial component $F_\rho$ of the force. A force that is purely tangential ($F_\rho=0$) produces no yaw torque in this leading-order model.\\

To calculate the torque as a function of the angular velocity, Newton's second law in the radial direction is used
\begin{equation}
m(-\rho\dot{\varphi}^2) = F_\rho + (T_f)_\rho,
\end{equation}
where $(T_f)_\rho$ is the radial component of the friction resultant. In the translation-dominated regime, the friction resultant is opposite the direction of motion, hence tangential, so $(T_f)_\rho\approx 0$. Thus,
\begin{equation}
\boxed{
F_\rho \approx -m\rho\dot{\varphi}^2 = -\frac{m v^2}{\rho},
}
\end{equation}
with $v=\norm{\vv}=\rho\dot{\varphi}$ under $\dot{\varphi}>0$.

Substituting $F_\rho\approx -m\rho\dot{\varphi}^2$ into $\tau_z=-\mu\Delta z F_\rho$ gives
\begin{equation}
\boxed{
\tau_z \approx \mu\,\Delta z\,m\rho\dot{\varphi}^2.
}
\end{equation}
If the force is applied below the CM, then $\Delta z<0$, so the torque tends to drive yaw in the opposite sense to the orbital motion (under the $\dot{\varphi}>0$ convention).

\subsection{\label{subsec:yaw_dynamics}Yaw dynamics: integrating Euler's equation}

Euler's equation is
\[
I_z\dot{\omega}=\tau_z.
\]
Using the circular-motion estimate above,
\begin{equation}
I_z\dot{\omega} \approx \mu\,\Delta z\,m\rho\dot{\varphi}^2.
\end{equation}
If $\dot{\varphi}$ is maintained approximately constant, the right-hand side is approximately constant, implying
\begin{equation}
\omega(t) \approx \omega(0) + \frac{\mu\,\Delta z\,m\rho\dot{\varphi}^2}{I_z}\,t.
\end{equation}
This predicts an approximately linear growth of yaw rate at early times, until higher-order effects become important.

% -----------------------------------------------------------
\section{\label{sec:numerical}Numerical Analysis of the Finite Rotation Regime}

The previous analytical approximation relies on the assumption that $\omega R \ll v$. This holds true only for the initial moments of motion. As the torque acts over time, the angular velocity $\omega$ increases. Eventually, the rotational speed at the rim becomes comparable to the translational speed.

Mathematically, this introduces a significant nonlinearity. The friction direction vector depends on the ratio of rotation to translation. As this ratio changes, the friction force vectors no longer align in parallel, but form a complex vortex-like pattern on the contact ring. This geometric shifting reduces the efficiency of the yaw torque generation.

To characterize the full range of motion up to the steady state, we must abandon the linear approximation and numerically simulate the governing equations.

\subsection{\label{subsec:exact_numerical}Exact Numerical Formulation}
The exact magnitude of the slip velocity vector at any angle $\theta$ is given by:
\begin{equation}
    \|\bm{v}_{\mathrm{rel}}(\theta)\| = \sqrt{v^2 + (\omega R)^2 - 2v\omega R \sin\theta}.
\end{equation}
The torque $\tau_z$ is the integral of the moment of the friction forces. Substituting the exact slip magnitude into the general torque equation yields:
\begin{equation}
    \tau_z(\omega) = \int_0^{2\pi} -\mu \lambda(\theta) \frac{R(v\sin\theta - \omega R)}{\sqrt{v^2 + (\omega R)^2 - 2v\omega R \sin\theta}} \, d\theta.
\end{equation}

Lastly, the time evolution of the system is governed by Euler's equation:
\begin{equation}
    I_z \frac{d\omega}{dt} = \tau_z(\omega).
\end{equation}

\subsection{\label{subsec:results_discussion}Results and Discussion}
\begin{figure*}[t]
    \centering
    % Panel A: Torque vs Omega
    \begin{subfigure}[b]{0.45\textwidth}
        \centering
        \includegraphics[width=\textwidth]{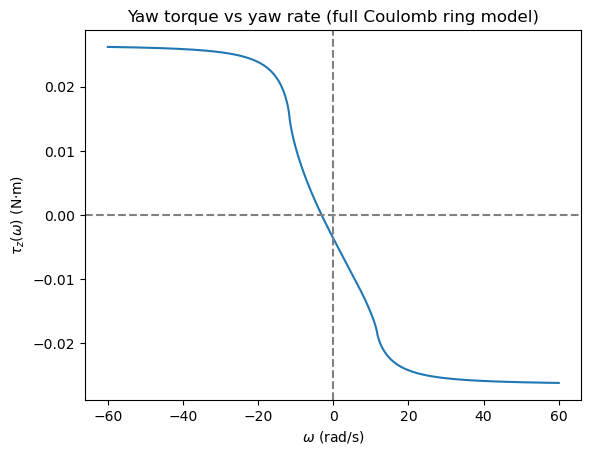}
        \caption{Yaw torque as a function of angular velocity.}
        \label{fig:torque_decay}
    \end{subfigure}
    \hfill
    % Panel B: Omega vs Time
    \begin{subfigure}[b]{0.445\textwidth}
        \centering
        \includegraphics[width=\textwidth]{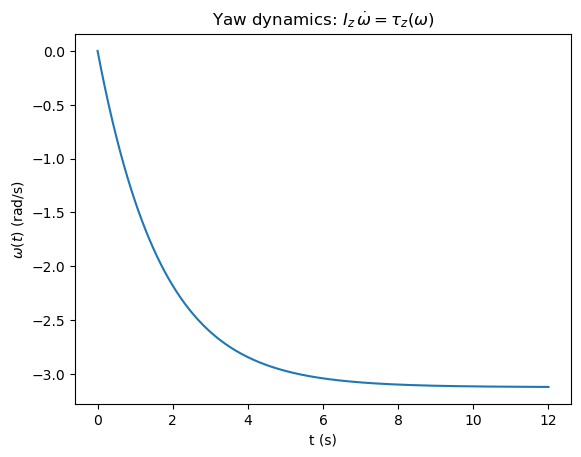}
        \caption{Time evolution of the yaw angular velocity.}
        \label{fig:time_evolution}
    \end{subfigure}
    
    \caption{\textbf{Dynamics of the Finite Rotation Regime.} (a) Numerical evaluation of the yaw torque $\tau_z$ as a function of angular velocity $\omega$. The zero-crossing point identifies the terminal steady-state velocity (approximately \SI{-3.1}{rad/s} in this configuration). (b) The resulting time evolution of the angular velocity. Starting from rest, the glass accelerates in the negative direction, asymptotically approaching the terminal velocity as the driving torque is balanced by the rotational friction effects. Simulation parameters: $m=\SI{0.30}{kg}$, $R=\SI{30}{mm}$, $\rho=\SI{0.15}{m}$, $v=\SI{0.35}{m/s}$, $\Delta z=\SI{-50}{mm}$ (force applied below the CM), and $\mu=0.30$.}

    \label{fig:combined_dynamics}
\end{figure*}

To solve this system, we employ a discrete time-stepping (forward) Euler algorithm. At each time step, the instantaneous torque $\tau_z(\omega_i)$ is calculated by numerical quadrature of the contact integral, and Euler’s equation is then used to update the angular velocity via $\omega_{i+1}=\omega_i+\big(\tau_z(\omega_i)/I_z\big)\Delta t$. Iterating this procedure yields the full time evolution $\omega(t)$, allowing us to capture the transient behavior and the eventual approach to a steady-state yaw rate.
The results of the numerical integration are presented in Figure \ref{fig:combined_dynamics}.

% Two-panel figure: in two-column mode, consider figure* if it looks cramped.

Figure \ref{fig:torque_decay} illustrates the non-linear relationship between the yaw torque and the angular velocity. The non-zero torque at $\omega = 0$ represents the initial driving torque derived in Section 5, which initiates the rotation. As the magnitude of the angular velocity increases, the torque magnitude decreases because the rotational component of the friction vector begins to counteract the translational asymmetry. Eventually, the system reaches a terminal velocity where the net torque vanishes, indicated by the zero-crossing at approximately \SI{-3.1}{rad/s}. Figure \ref{fig:time_evolution} validates this behavior, showing the angular velocity starting from rest and asymptotically approaching this steady-state value as the driving torque and rotational friction reach equilibrium.

Taken together, the simulation confirms two key points. First, it proves that the pressure redistribution mechanism generates a finite yaw torque at $\omega=0$, thus providing a dynamical rotation without the need of any externally applied yaw moment. Second, it demonstrates that the complete Coulomb ring model predicts a stable steady state: as $\omega$ increases, the contact kinematics modify the slip field so that the net torque decreases and eventually cancels, producing a well-defined steady state yaw rate instead of an unbounded increase in spin rate. This numerical result therefore accurately correlates the initial analytical regime and the experimentally observed saturation of the self-rotation.

% -----------------------------------------------------------
\subsection{\label{sec:limitations}Model assumptions and limitations}

The model developed above is intentionally minimal. Its purpose is to isolate the mechanism by which a redistributed normal reaction produces a yaw torque, rather than to provide a complete theory of glass--table contact. Several assumptions should therefore be kept in mind.

First, the contact between the glass and the table has been represented as an annular ring of radius $R$. Real glass bases are not perfectly rigid rings: the actual contact occurs through microscopic asperities and may depend on surface roughness, small geometrical imperfections and local compliance. This is a standard limitation of macroscopic Coulomb-friction models \cite{bowden1950friction,greenwood1966contact,johnson1985contact}.

Second, the normal load distribution was approximated by its lowest non-trivial Fourier component,
\[
\lambda(\theta)=\frac{N}{2\pi}+A\cos\theta+B\sin\theta.
\]
This first-harmonic form is sufficient to reproduce the total normal force and the in-plane moment required for quasi-static tilt equilibrium. Higher harmonics would change the local pressure distribution but would not alter the leading-order force and moment balance. The model is therefore best understood as a lowest-order closure rather than as a detailed contact-pressure calculation.

Third, the model assumes that the whole annular base remains in contact with the table. This imposes the condition
\[
\lambda(\theta)\geq 0
\quad \text{for all } \theta.
\]
With the first-harmonic distribution, this requires
\[
\frac{N}{2\pi}
\geq
\sqrt{A^2+B^2}.
\]
Using
\[
A=\frac{M_{h,y}}{\pi R},
\qquad
B=-\frac{M_{h,x}}{\pi R},
\]
one obtains the approximate no-lift-off condition
\[
|\mathbf{M}_h|\leq \frac{NR}{2}.
\]
If this condition is violated, part of the base may unload and the full-ring model is no longer valid.

Fourth, kinetic Coulomb friction has been used locally at each point of the contact ring. Real friction may include static friction, stick--slip effects, velocity dependence, surface contamination and rolling or rocking losses. These effects may change the quantitative value of the angular acceleration and terminal yaw rate, but they do not remove the central mechanism: an asymmetric normal load produces an asymmetric friction field and therefore a yaw torque.

Finally, the orbital motion has been treated as imposed. In the actual hand-driven experiment, the applied force, the orbital speed and the radius are only approximately controlled. For this reason, the model should be interpreted primarily as a physical explanation and as a semi-quantitative teaching model. A closer experimental comparison would require direct tracking of the centre-of-mass trajectory and the yaw angle throughout the motion.

% -----------------------------------------------------------
\section{\label{sec:conclusion}Conclusion}

We have analyzed a simple tabletop phenomenon in which a wine glass, when forced to follow a circular path, develops a spontaneous yaw rotation even when the applied hand force does not directly cause a yaw torque. The central result is that the yaw originates from the distributed pressure at the table contact: a vertical offset between the force application point and the center of mass produces an overturning moment that redistributes the normal load around the base ring. Because local kinetic friction scales with the local normal load, this redistribution breaks the symmetry of the friction field and generates a net torque about the vertical axis.

In the translation-dominated regime, we obtained a closed-form expression for the yaw torque and its dependence on the force offset, friction coefficient, and orbital radius. The analysis clarifies why applying the force near the center of mass height produces little or no yaw, while applying it lower on the stem yields a self-rotation. Beyond this approximation, a full Coulomb ring model requires numerical evaluation of the torque, and integrating Euler's equation shows that the yaw rate does not grow unbounded but approaches a finite steady-state value as the rotational contribution to slip progressively reduces the driving asymmetry.

Together, the analytical and numerical results provide a compact mechanical explanation and quantitative predictions for this reproducible at-home experiment. More broadly, the example illustrates how small changes in force application can reorganize contact pressures and convert translation into rotation through distributed friction.

\section*{Author Declarations}
The author has no conflicts to disclose.

\end{document}